\documentclass[fleqn,12pt,twoside]{article}
\usepackage{espcrc1}

\usepackage{graphicx}
\usepackage[figuresright]{rotating}

\title{Photoproduction of Pseudoscalar Mesons}

\author{R. A. Arndt\address[GW]{The Center for Nuclear Studies, 
              Department of Physics, \\
              The George Washington University, Washington, 
              D.C. 20052},
        W. J. Briscoe\addressmark[GW],
        G. V. O'Rielly\address[UMAS]{Department of Physics,
              University of Massachusetts 
              Dartmouth, M.A. 02747},
        I. I. Strakovsky\addressmark[GW],
        R. L. Workman\addressmark[GW]}
       
\begin{document}

\maketitle

\begin{abstract}
Experiments that study the photoproduction of
pseudoscalar mesons; pions, etas and kaons, have 
the potential to increase our knowledge of
baryon and hyperon resonance properties.  Recent 
experiments at JLab, Mainz, GRAAL, and Bonn are 
beginning to produce results in the form of 
polarization and asymmetry measurements and 
determinations of the differential and integrated 
cross sections.  These new data are essential
to the performance of Partial-Wave Analyses that 
are less model dependent and coupled-channels
calculations that incorporate unitarity dynamically,
combining hadronic reaction channels together with
electromagnetic processes.  This approach is 
necessary to extract resonance properties and may 
lead to the identification of missing, but predicted, 
resonances.  Some recent experimental and 
phenomenological results for single and double
pseudoscalar meson photoproduction are discussed. \\
\end{abstract}

%%%%%%%%%%%%%%%%%%%%%%%%%%%%%%%%%%%%%%%%%%%%%%%%%%%%%%%
Studies of the baryon spectrum have generally been 
data-base driven.  While most states were first 
detected through hadronic scattering (mainly 
pion-nucleon), present research is focused on 
resonance excitation by real and virtual photon 
probes. The photo- and electroproduction of 
pseudoscalar mesons has been studied most
intensively for a number of practical reason. 
First, in the case of pion production, there is a 
clear and direct connection to the extensive 
results from elastic pion-nucleon scattering.  Here 
one benefits from knowing, in advance, the 
resonances which couple most strongly to the final 
state.  Second, in general, pseudoscalar-meson 
production is relatively easy to analyze.  Many 
additional amplitudes contribute to the production 
of vector mesons, for example. However, if one is 
looking for ``missing resonances," states {\it not} 
found in the elastic scattering or electromagnetic 
production of single pseudoscalar mesons, the 
search must be extended to processes having other 
final states (such as two-meson productions).

It has become clear over the last decade that any 
attempt at an unambiguous description of resonances 
requires a comprehensive approach to the analysis 
of all contributing reactions.  In order to be able 
to perform such coupled-channel analyses, one needs
easy access to all related sets of experimental 
data, in addition to a theoretical and calculational 
framework for a common description.  The 
availability of partial-wave amplitudes greatly 
simplifies numerical aspects of coupled-channel 
analysis.  The number of fitted partial-wave 
amplitudes associated with a dataset may be smaller 
than the count of individual data by orders of 
magnitude, and can account for issues associated 
with statistical and systematic errors and the 
rejection of inconsistent measurements.

\begin{figure}[htb]
\begin{minipage}[c]{70mm}
\includegraphics[angle=90,width=17pc]{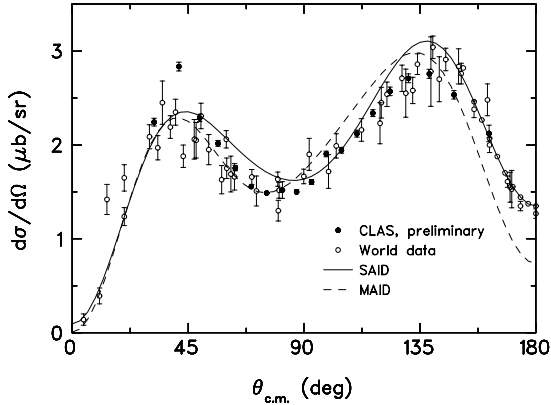}
\end{minipage}
\hspace{\fill}
\begin{minipage}[c]{80mm}
         \caption{Preliminary differential cross 
         section data in the p($\gamma$, p)$\pi^0$ 
         reaction at E$_{\gamma}$ = 1000~MeV from 
         the CLAS $g1c$ run group (solid circles)~
         \protect\cite{wood}.  Data are compared 
         to the world data (open circles), along 
         with SM02~\protect\cite{sm02} solution 
         (solid line) and MAID~\protect\cite{maid} 
         (dashed line) predictions.}
\label{fig:fig1}
\end{minipage}
\end{figure}

In Fig.~\ref{fig:fig1}, we compare the SAID and 
MAID fits to existing data and preliminary CLAS 
data for single-pion photoproduction.  Note that 
the variation in fits to existing data would be 
unacceptable if only CLAS data were retained.  
Differences, usually minor in the unpolarized 
cross sections, are magnified in predictions of 
single and double polarization observables.
%%%%%%%%%%%%%%%%%%%%%%%%%%%%%%%%%%%%%%%%%%%%%%%%%%%%%%%

\begin{figure}[htb]
\begin{minipage}[t]{70mm}
\includegraphics[angle=90,width=17pc]{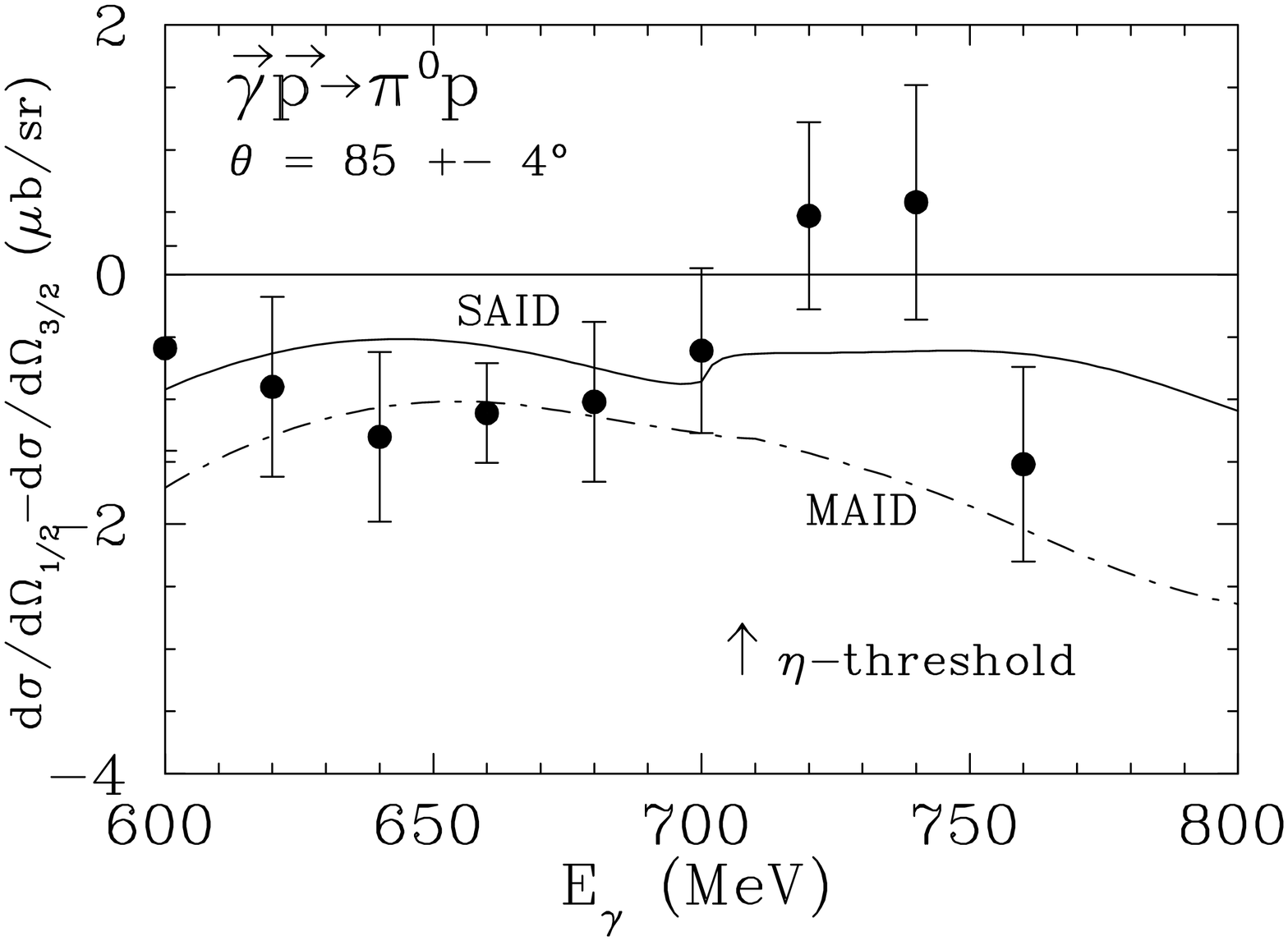}
\end{minipage}
\hspace{\fill}
\begin{minipage}[t]{75mm}
\includegraphics[angle=90,width=17pc]{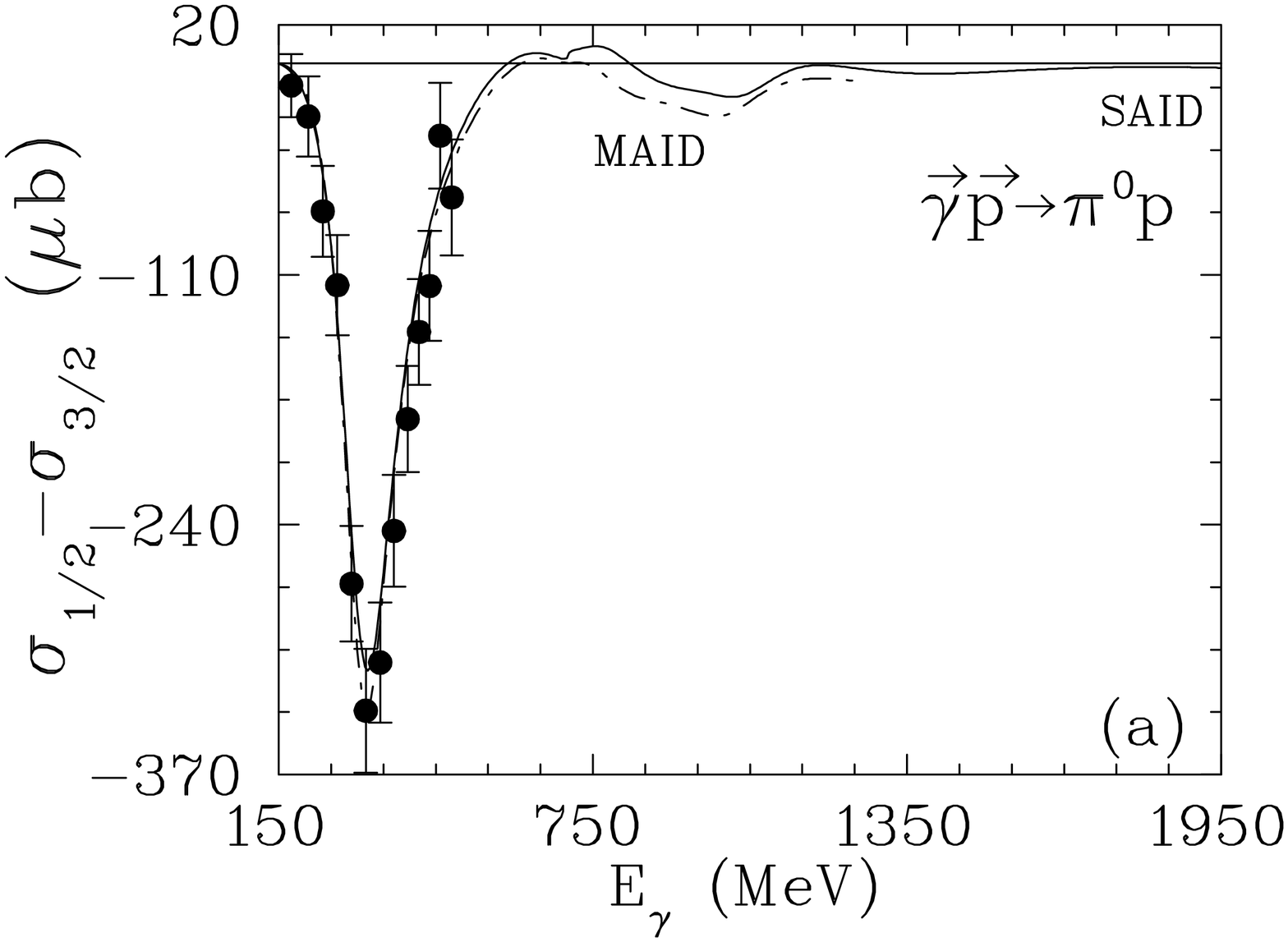}
\end{minipage}
\caption{Difference of the (a) differential and (b) 
         total cross sections for the helicity 
         states 1/2 and 3/2 for $\vec{\gamma}\vec{p}
         \to\pi^0p$.  The solid (dash-dotted) line 
         represents the SM02~\protect\cite{sm02} 
         (MAID2001~\protect\cite{maid}) solution.  
         Experimental data are from Mainz~
         \protect\cite{dx13,ah00}.}
\label{fig:fig2}
\end{figure}

In Fig.~\ref{fig:fig2}, we compare SAID and MAID 
results for the differential and total cross section 
(polarized beam-target) contributing to the the GDH 
sum rule.  The inclusion of this and other precise 
polarization data has led a number of groups to 
propose changes to supposedly well-known resonance 
states.  Note that the single-pion contribution to 
the sum rule is negligible by 2~GeV.
%%%%%%%%%%%%%%%%%%%%%%%%%%%%%%%%%%%%%%%%%%%%%%%%%%%%%%%

\begin{figure}[htb]
\begin{minipage}[c]{70mm}
\includegraphics[angle=0,width=17pc]{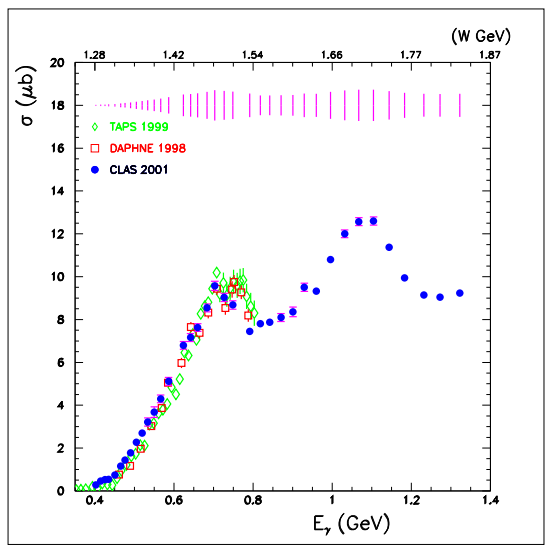}
\end{minipage}
\hspace{\fill}
\begin{minipage}[c]{85mm}
         \caption{Total cross section for the
         $\gamma p\to p\pi^0\pi^0$ reaction.
         Preliminary CLAS data~\protect\cite{sasha} 
         are compared to the world data (open 
         circles) obtained with the DAPHNE'98 and
         TAPS'99 detectors.  CLAS data are plotted
         with statistical errors.  The band at the
         top of the figure delineates the range of
         systematic uncertainties.}
\label{fig:fig3}
\end{minipage}
\end{figure}

In Fig.~\ref{fig:fig3}, we show results for double-pion 
photoproduction, which displays structures which have 
been attributed to states not strongly seen in 
single-pion production.  Here, however, interpretation
is more difficult, again motivating a multi-channel 
treatment.
%%%%%%%%%%%%%%%%%%%%%%%%%%%%%%%%%%%%%%%%%%%%%%%%%%%%%%%

\begin{figure}[htb]
\begin{minipage}[c]{60mm}
\includegraphics[angle=0,width=17pc]{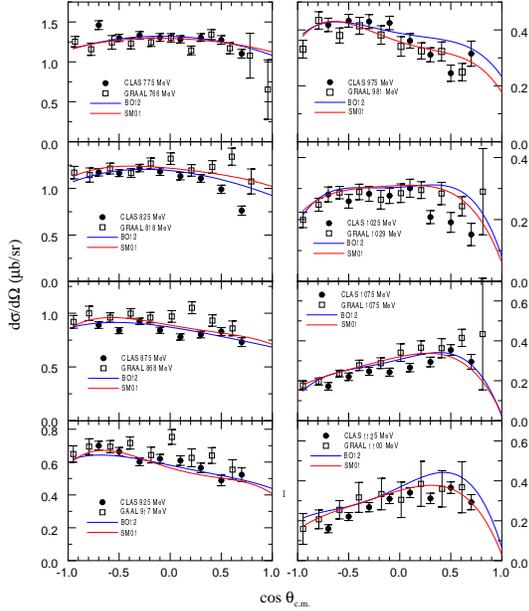}
\end{minipage}
\hspace{\fill}
\begin{minipage}[c]{85mm}
         \caption{CLAS $\gamma p\to\eta p$ 
         differential cross sections.  CLAS data 
         (solid circles)~\protect\cite{zhen} are 
         compared to the GRAAL data (open squares)~
         \protect\cite{graal} along with SM01 
         solution with CLAS data included, and 
         BO12 solution with CLAS data excluded~
         \protect\cite{gpep}.}
\label{fig:fig4}
\end{minipage}
\end{figure}

In Fig.~\ref{fig:fig4}, two GW fits to the S-wave 
dominated eta-photoproduction differential cross 
sections are displayed.  This reaction has the
potential to give cleaner information on the N(1535) 
resonance, the pion channel being masked by a sharp 
cusp effect (due to the eta-nucleon threshold).  Note 
that both fits, and also a fit from the Mainz group, 
tends to drop off rapidly at very forward angles.  
An interesting preliminary set of data from GRAAL 
suggests a flatter extrapolation to the forward 
direction, which would necessitate a re-examination 
of the underlying assumptions in most fits (a smoother 
behaviour appears to be predicted by the model of Li 
and Saghai.)
%%%%%%%%%%%%%%%%%%%%%%%%%%%%%%%%%%%%%%%%%%%%%%%%%%%%%%%

Fits to pion electroproduction and kaon photoproduction 
have also been completed, and a multi-channel K-matrix 
approach to eta photoproduction is being explored. 
Further fits and an expansion of our on-line database 
is expected in the near future.

%%%%%%%%%%%%%%%%%%%%%%%%%%%%%%%%%%%%%%%%%%%%%%%%%%%%%%%
\section*{Acknowledgments}
This work has been supported by U.S. Department of 
Energy grant DE-FG02-99-ER41110.  Additional support 
has been provided by Jefferson Lab and the 
Southeastern Universities Research Association through 
the D.O.E. contract DE-AC05-84ER40150.

%%%%%%%%%%%%%%%%%%%%%%%%%%%%%%%%%%%%%%%%%%%%%%%%%%%%%%%

\end{document}